\documentclass[sn-mathphys-num]{sn-jnl}

\usepackage{graphicx}%
\usepackage{multirow}%
\usepackage{amsmath,amssymb,amsfonts}%
\usepackage{amsthm}%
\usepackage{mathrsfs}%
\usepackage[title]{appendix}%
\usepackage{textcomp}%
\usepackage{manyfoot}%
\usepackage{booktabs}%
\usepackage{algorithm}%
\usepackage{algorithmicx}%
\usepackage{algpseudocode}%
\usepackage{listings}%
\usepackage[table]{xcolor}

\usepackage{subcaption}

\newcommand{\gptfour}{{{GPT-4}}}
\newcommand{\gptfouro}{{GPT-4o}}
\newcommand{\gptfouromini}{{GPT-4o-Mini}}
\newcommand{\oonepreview}{{O1-Preview}}
\newcommand{\oonemini}{{O1-Mini}}
\newcommand{\claude}{{Claude-3.5-Sonnet}}
\newcommand{\geminipro}{{Gemini-1.5-Pro-002}}
\newcommand{\geminiflash}{{Gemini-1.5-Flash-002}}
\newcommand{\llamathreeone}{{Llama-3.1-405B-Instruct}}
\newcommand{\nemotron}{{Llama-3.1-Nemotron-70B-Instruct}}
\newcommand{\llamathreetwo}{{Llama-3.2-90B-Vision-Instruct}}
\newcommand{\qwen}{{Qwen2.5-72B-Instruct}}
\newcommand{\mistral}{{Mistral-Large-2}}
\newcommand{\openmixtral}{{Mixtral-8x22B}}
\newcommand{\wizard}{{WizardLM-2-8x22B}}
\newcommand{\codestral}{{Codestral}}

\begin{document}

\title{Evaluating Large Language Models on Business Process Modeling: Framework, Benchmark, and Self-Improvement Analysis}


\author*[1,2]{\fnm{Humam} \sur{Kourani}}\email{humam.kourani@fit.fraunhofer.de}

\author[2]{\fnm{Alessandro} \sur{Berti}}\email{a.berti@pads.rwth-aachen.de}

\author[2,3]{\fnm{Daniel} \sur{Schuster}}\email{schuster@pads.rwth-aachen.de}

\author[1,2]{\fnm{Wil M.P.} \sur{van der Aalst}}\email{wvdaalst@pads.rwth-aachen.de}

\affil*[1]{\orgname{Fraunhofer Institute for Applied Information Technology FIT}, \orgaddress{\street{Schloss Birlinghoven}, \city{Sankt Augustin}, \postcode{53757}, \country{Germany}}}

\affil[2]{\orgname{RWTH Aachen University}, \orgaddress{\street{Ahornstraße 55}, \city{Aachen}, \postcode{52074}, \country{Germany}}}

\affil[3]{\orgname{Process Intelligence Solutions}, \orgaddress{\street{Kurfürstenstraße 5}, \city{Aachen}, \postcode{52066}, \country{Germany}}}

\abstract{Large Language Models (LLMs) are rapidly transforming various fields, and their potential in Business Process Management (BPM) is substantial. This paper assesses the capabilities of LLMs on business process modeling using a framework for automating this task, a comprehensive benchmark, and an analysis of LLM self-improvement strategies. We present a comprehensive evaluation of 16 state-of-the-art LLMs from major AI vendors using a custom-designed benchmark of 20 diverse business processes. Our analysis highlights significant performance variations across LLMs and reveals a positive correlation between efficient error handling and the quality of generated models. It also shows consistent performance trends within similar LLM groups. Furthermore, we investigate LLM self-improvement techniques, encompassing self-evaluation, input optimization, and output optimization. Our findings indicate that output optimization, in particular, offers promising potential for enhancing quality, especially in models with initially lower performance. Our contributions provide insights for leveraging LLMs in BPM, paving the way for more advanced and automated process modeling techniques.
}




\maketitle


\section{Introduction}
\label{sec:introduction}

Process modeling is a crucial component of Business Process Management (BPM), acting as a comprehensive toolkit for understanding, documenting, analyzing, and optimizing intricate business operations. It encompasses various forms – from textual descriptions to visual diagrams and executable models – thereby providing a multi-dimensional approach to capturing the nuances of organizational processes.

Business process modeling integrates several key perspectives, each focusing on distinct aspects of processes. These include the \emph{control-flow perspective}, which maps out the sequence of activities and their interdependencies; the \emph{data perspective}, which deals with the creation, manipulation, and usage of data throughout the process; the \emph{resource perspective}, which identifies the human and system resources required for process execution; and the \emph{operational perspective}, which outlines the rules and execution semantics governing the process. Our focus in this paper is primarily on enhancing the control-flow perspective because it forms the foundational structure that supports the integration of data, resources, and operational aspects into a process model.

Traditionally, business process modeling requires considerable manual effort and a deep understanding of sophisticated process modeling languages such as BPMN (Business Process Model and Notation) \cite{DBLP:books/el/15/RosingWCM15} and Petri nets \cite{DBLP:journals/topnoc/HeeSW13a}. Moreover, maintaining these process models to reflect changes in business operations is an ongoing challenge, presenting significant obstacles for individuals lacking expertise in these languages, thus highlighting the need for more streamlined methodologies in process modeling.

The emergence of Large Language Models (LLMs) such as GPT-4 \cite{DBLP:journals/corr/abs-2303-08774} and Gemini \cite{DBLP:journals/corr/abs-2312-11805} offers a promising avenue for enhancing the efficiency and accessibility of process modeling. Trained on vast and varied datasets, these models are skilled in a range of tasks from generating coherent and contextually relevant text to solving complex problems and producing executable code \cite{ijcai2021p612,DBLP:conf/hpec/VidanF23,DBLP:conf/iclr/ZhouMHPPCB23}. Their capability to process and interpret complex textual inputs in natural language positions LLMs as particularly suitable for tasks like process modeling that require the generation and refinement of structured outputs from textual descriptions.

Our previously introduced framework \cite{DBLP:conf/bpmds/KouraniB0A24} leverages LLMs to automate the generation and refinement of process models from textual descriptions. It employs sophisticated techniques in prompt engineering, error handling, and code generation, transforming detailed process descriptions into process models. This framework utilizes the Partially Ordered Workflow Language (POWL) \cite{DBLP:conf/bpm/KouraniZ23} as an intermediate representation due the quality guarantees it provides, particularly in ensuring \emph{soundness} \cite{DBLP:conf/bpm/KouraniZ23}, and the possibility to export POWL models in standard notations such as BPMN and Petri nets. Preliminary results in \cite{DBLP:conf/bpmds/KouraniB0A24} demonstrated the practicality and effectiveness of this framework, showcasing its clear advantage over alternative solutions due to the usage of POWL to ensure soundness. 

This paper extends the work presented in \cite{DBLP:conf/bpmds/KouraniB0A24} by introducing two key new contributions: a comprehensive benchmarking analysis and an investigation into LLM self-improvement strategies. To rigorously evaluate the performance of LLMs within our framework for automated process modeling, we design a comprehensive benchmark consisting of 20 diverse business processes, each paired with a ground-truth process model and a simulated event log. This setup allows for an automated yet qualitative assessment of the generated process models using conformance checking techniques \cite{DBLP:conf/s-bpm-one/DunzerSMB19}. We evaluate 16 state-of-the-art LLMs from various AI vendors, such as Google (\url{https://ai.google.dev/}), OpenAI (\url{https://openai.com/}), Anthropic (\url{https://www.anthropic.com/}), Meta (\url{https://ai.meta.com/}), and Mistral AI (\url{https://mistral.ai/}). This benchmark allows us to assess the diverse capabilities of LLMs, including natural language understanding, code generation, adherence to instructions, and error correction.

Beyond benchmarking, we explore the potential of LLM self-improvement techniques to further enhance the quality of generated process models. We investigate three specific strategies: self-evaluation, input optimization, and output optimization. By leveraging these techniques, we aim to assess whether LLMs can autonomously refine their performance, leading to more accurate and reliable process models.

The structure of this paper is as follows. First, related work is discussed in \autoref{sec:relatedWork}. \autoref{sec:methodology} provides a detailed overview of our LLM-based process modeling framework. \autoref{sec:tool} introduces the ProMoAI tool, which supports our framework. \autoref{sec:ev} presents the benchmarking analysis, including the experimental setup, evaluation metrics, and a discussion of the results. \autoref{sec:ev2} investigates the LLM self-improvement strategies and analyzes their impact on model quality. Finally, \autoref{sec:conclusion} concludes the paper.

\section{Related Work}
\label{sec:relatedWork}

An overview of various methods for extracting process information from textual content is presented in \cite{DBLP:conf/aiia/BellanDG20}. The study in \cite{DBLP:journals/jucs/GoncalvesSB11} utilizes Natural Language Processing (NLP) and text mining techniques to derive process models directly from text, while \cite{DBLP:conf/caise/FriedrichMP11} combines NLP with computational linguistics to generate BPMN models. The approach in \cite{DBLP:journals/jksucis/SholiqSA22} applies NLP to extract structured relationship representations, referred to as \emph{fact types}, from textual data, which are then converted into BPMN components. The BPMN Sketch Miner, as detailed in \cite{DBLP:conf/models/IvanchikjSP20}, uses process mining \cite{DBLP:books/daglib/0027363} to produce BPMN models from text described in a \emph{domain-specific language}. Commercial solutions are also adopting AI for process modeling; for example, Process Talks (\url{https://processtalks.com}) offers an AI-driven platform for generating process models from textual descriptions.

Recent studies have explored the integration of LLMs in BPM, investigating their potential applications and challenges. Several works \cite{DBLP:conf/bpmds/BuschRSL23,DBLP:conf/bpm/VidgofBM23} delve into how LLMs can be employed for BPM and process mining tasks. The papers \cite{DBLP:journals/corr/abs-2408-17316} and \cite{DBLP:journals/corr/abs-2408-08892} explore the application of LLMs in process discovery and process querying, respectively. The research in \cite{DBLP:journals/aai/ChenL22b} employs BERT \cite{DBLP:conf/naacl/DevlinCLT19} to classify and analyze process execution logs, aiming to enhance process monitoring and anomaly detection. The limitations of using GPT-4 for conceptual modeling are discussed in \cite{20.500.12116/43782}. In \cite{DBLP:conf/bpm/KlievtsovaBKMR23,DBLP:journals/corr/abs-2407-17478,DBLP:conf/bpm/Fontenla-SecoWG23}, methods for generating process models through dialogue-based interactions and chatbots are proposed. The paper \cite{DBLP:conf/bpm/GrohsAER23} showcases the ability of LLMs to translate textual descriptions into procedural and declarative process model constraints. Finally, \cite{DBLP:journals/emisaij/FillFK23} investigates the broader impacts of LLMs in conceptual modeling, proposing potential applications that extend beyond traditional BPM tasks.

Several benchmarks for evaluating LLMs on process mining and BPM tasks are proposed. The benchmark in \cite{DBLP:journals/corr/abs-2407-13244} assesses LLMs across a spectrum of process mining tasks, utilizing self-evaluation by LLMs to judge the quality of results. This approach contrasts with the benchmark proposed in this paper, which uses process descriptions aligned with ground truth models, allowing for a more informed and objective assessment of model quality. Further studies such as \cite{DBLP:journals/corr/abs-2406-05506,DBLP:journals/corr/abs-2401-12846} propose benchmarks that focus on causal reasoning and the explanation of decision points within business processes. Additionally, \cite{DBLP:conf/icpm/RebmannSGA24} introduces benchmarks for semantic-aware process mining tasks like semantic anomaly detection and next activity prediction. 

This paper extends our previous work \cite{DBLP:conf/bpmds/KouraniB0A24} by incorporating the following contributions. The evaluation of the proposed LLM-based process modeling framework has been significantly broadened to include a larger and more diverse set of processes, as well as a wider range of state-of-the-art LLMs. Additionally, a robust qualitative assessment methodology has been implemented, employing conformance checking \cite{DBLP:conf/s-bpm-one/DunzerSMB19} against ground truth event logs to enable a more rigorous evaluation of the generated process models. Furthermore, we explore LLM self-improvement techniques to assess whether LLMs can autonomously refine their performance within our framework.


\section{LLM-Based Process Modeling Framework}
\label{sec:methodology}
In this section, we present a detailed overview of our framework, which harnesses the capabilities of LLMs to generate and refine process models based on natural language process descriptions. 

\subsection{Framework Overview}

\begin{figure}[!t]
    \centering        
    \includegraphics[width=\textwidth]{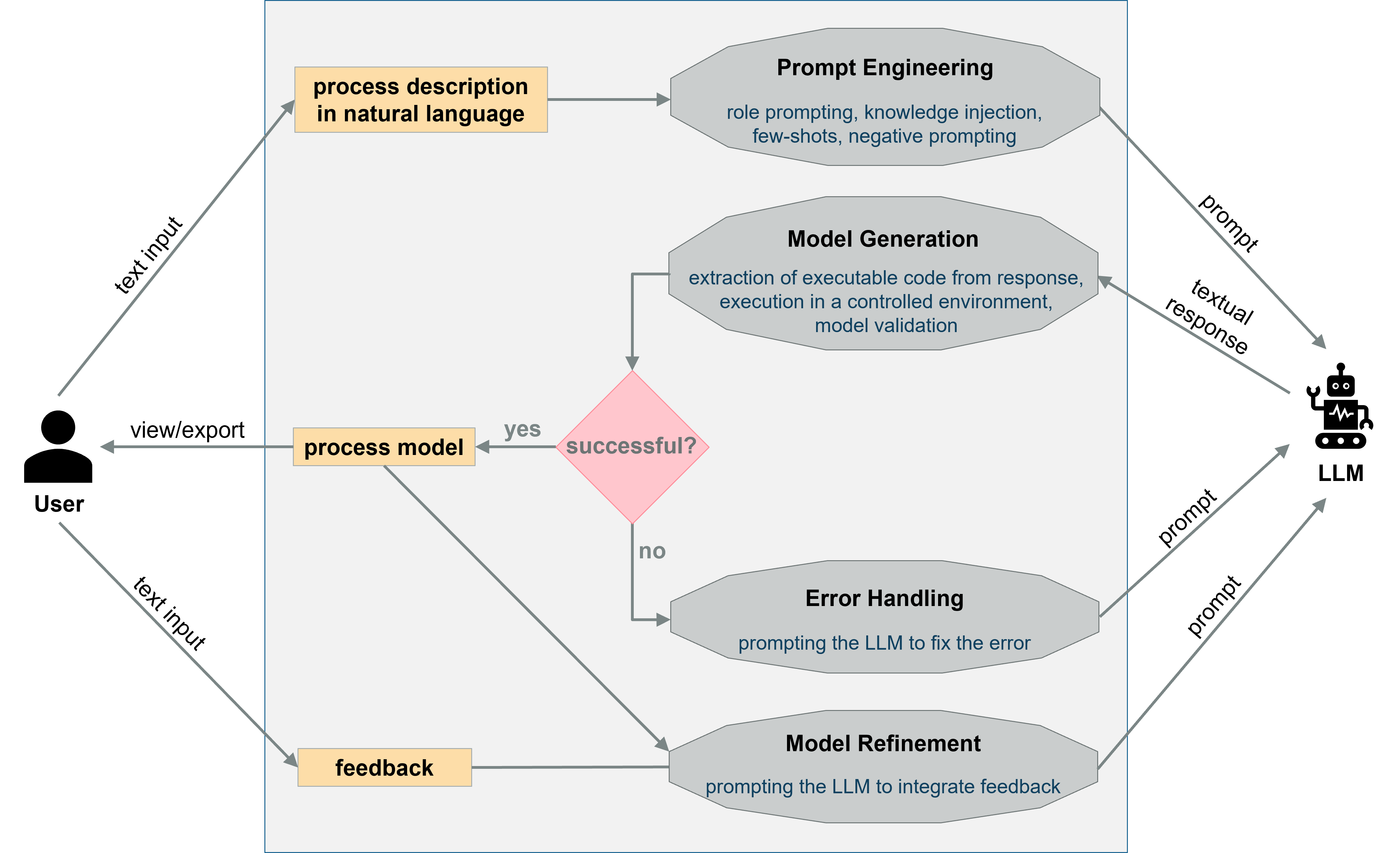}
    \caption{LLM-based process modeling framework.}
     \label{fig:architecture}
 \end{figure}

\autoref{fig:architecture} offers a high-level view of our proposed framework. 
The framework begins by having users provide a textual description of a process in natural language. After receiving the process description, additional information is integrated to create a comprehensive prompt (the prompt engineering strategies are discussed in \autoref{sec:prompt}). This prompt is carefully crafted to instruct the LLM in generating executable code that can then be used to create process models (the selection of the modeling language is discussed in \autoref{sec:powl}). A set of functions designed specifically for process model creation aids in this code generation. Once the prompt is prepared, it is sent to the LLM. Our framework is not dependent on any specific LLM and can function with any advanced LLM that supports a large context window and code generation. After receiving the LLM's response, we extract the generated code and attempt to execute it (details in \autoref{sec:gen}). 

In case errors are encountered during code extraction or execution, an error-handling mechanism is activated, sending a refined prompt back to the LLM to address the issue (discussed in \autoref{sec:error}). Upon successful execution and process model creation, users can view or export the model using established process modeling notations such as BPMN and Petri nets. Furthermore, the framework allows users to provide feedback on the generated model, which can then be incorporated to further refine the model, enabling continuous improvement.

\subsection{Process Representation}\label{sec:powl}

To explain the various stages of our framework, we implement an instance that uses the Partially Ordered Workflow Language (POWL) \cite{DBLP:conf/bpm/KouraniZ23} for intermediate process representation. The framework's core principles allow for integration with other modeling languages depending on process modeling requirements. This section highlights the reasons behind our choice of POWL.

Our objective is to generate process models using common notations that professionals in the business process management and process mining fields are familiar with, such as BPMN and Petri nets. However, these notations can lead to quality issues, as it is possible to generate models with unreachable parts, for instance. To address this, the concept of \emph{soundness} is introduced, and many automated process model discovery methods rely on languages that ensure soundness (e.g., \cite{DBLP:conf/icpm/KouraniSA23,DBLP:series/lnbip/Leemans22}). POWL is a partially ordered graph extended with control-flow operators to model choices and loops, representing a subclass of Petri nets that allows for the hierarchical combination of sub-models to form larger models. POWL models can be automatically transformed into Petri nets or BPMN models as described in \cite{DBLP:conf/bpm/KouraniZ23}.

POWL was selected as the intermediate representation for the following reasons:
\begin{itemize}
    \item \textit{Soundness Guarantees:} Unlike BPMN and Petri nets, POWL inherently guarantees soundness.
    \item \textit{Simplicity:} POWL's hierarchical structure simplifies model generation by enabling the recursive creation of sub-models, which are then combined into larger models. It also assumes tasks are parallel unless otherwise specified, reflecting the concurrent nature of many real-world processes. This assumption simplifies model generation since task order does not always need to be explicitly defined.
    \item \textit{Expressive Power:} While both POWL and process trees \cite{DBLP:series/lnbip/Leemans22} ensure soundness, POWL supports a wider range of process structures \cite{DBLP:conf/bpm/KouraniZ23}. It allows for the modeling of more complex dependencies while retaining the quality guarantees of hierarchical process modeling languages.
\end{itemize}

\subsection{Prompt Engineering}\label{sec:prompt}
This section outlines the prompt engineering strategies we use to guide the LLM in generating process models from natural language descriptions.

The key strategies we implemented within our framework are:
\begin{itemize}
    \item \textit{Role Prompting:} This approach involves assigning a specific role to the LLM to shape its behavior \cite{DBLP:journals/corr/abs-2305-14688}. We instruct the LLM to act as a process modeling expert, who is familiar with common process constructs. Additionally, we ask the LLM to act as a process owner, capable of filling in gaps in the process descriptions based on its expertise.
    
    \item \textit{Knowledge Injection:} This strategy refers to injecting specific knowledge that the LLM may not have encountered during its training \cite{DBLP:conf/esws/MartinoIT23}. We provide comprehensive knowledge about POWL, offering detailed insights into its hierarchical structure and the semantics of the different POWL components. Moreover, our framework leverages LLM capabilities in generating executable code \cite{DBLP:conf/hpec/VidanF23} by instructing the LLM to generate Python code that utilizes a predefined set of functions we designed for the safe generation of POWL models. We provide a detailed explanation of these predefined methods and how they can be used to generate POWL models. \autoref{lst:injection} illustrates the knowledge injected about POWL.
    
\begin{lstlisting}[caption={Knowledge injection on generating POWL models. Lines extending beyond the displayed text are abbreviated with ``...'' for compactness.}, label={lst:injection}, frame=single, float, floatplacement='!t', basicstyle=\footnotesize\ttfamily]
Use the following knowledge about the POWL modeling language: A POWL model is ...
Provide the Python code that recursively generates a POWL model. Save the final ...
Assume the class ModelGenerator is properly implemented and can be imported ...
ModelGenerator provides the functions described below:
 - activity(label) generates an activity. It takes 1 string argument, which is ...
 - xor(*args) takes n >= 2 arguments, which are the submodels. Use it to model ...
 - loop(do, redo) takes 2 arguments, which are the do and redo parts. Use it to ...
 - partial_order(dependencies) takes 1 argument, which is a list of tuples of ...
Note: for any powl model, you can call powl.copy() to create another instance ...
\end{lstlisting}

    \item \textit{Few-Shots Learning:} This technique involves providing the LLM with multiple example input-output pairs to train it on the task \cite{DBLP:conf/nips/BrownMRSKDNSSAA20}. For instance, \autoref{lst:few-shots} shows one of the few-shots examples we use for training the LLM to generate POWL models starting from process descriptions. Some of the process descriptions we use for implementing few-shots learning are adapted from \cite{DBLP:conf/bpm/BellanADGP22}.

\begin{lstlisting}[caption={One of the examples used for few-shots learning and negative prompting. Lines extending beyond the displayed text are abbreviated with ``...'' for compactness.}, label={lst:few-shots}, frame=single, float, floatplacement='!t', basicstyle=\footnotesize\ttfamily]
Process description for example 1:
A small company manufactures customized bicycles. Whenever the sales department ...

Process model for example 1:
```python
from utils.model_generation import ModelGenerator
gen = ModelGenerator()
create_process = gen.activity('Create process instance')
reject_order = gen.activity('Reject order')
accept_order = gen.activity('Accept order')
inform = gen.activity('Inform storehouse and engineering department')
process_part_list = gen.activity('Process part list')
check_part = gen.activity('Check required quantity of the part')
reserve = gen.activity('Reserve part')
back_order = gen.activity('Back-order part')
prepare_assembly = gen.activity('Prepare bicycle assembly')
assemble_bicycle = gen.activity('Assemble bicycle')
ship_bicycle = gen.activity('Ship bicycle')
finish_process = gen.activity('Finish process instance')

check_reserve = gen.xor(reserve, back_order)

single_part = gen.partial_order(dependencies=[(check_part, check_reserve)])
part_loop = gen.loop(do=single_part, redo=None)

accept_poset = gen.partial_order(
    dependencies=[(accept_order, inform),
                  (inform, process_part_list),
                  (inform, prepare_assembly),
                  (process_part_list, part_loop),
                  (part_loop, assemble_bicycle),
                  (prepare_assembly, assemble_bicycle),
                  (assemble_bicycle, ship_bicycle)])

choice_accept_reject = gen.xor(accept_poset, reject_order)

final_model = gen.partial_order(
    dependencies=[(create_process, choice_accept_reject),
                  (choice_accept_reject, finish_process)])
```

Common errors to avoid for example 1:
creating a local choice between 'reject_order' and 'accept_order' instead of ...
\end{lstlisting}
    
    \item \textit{Negative Prompting:} Negative prompting involves specifying what the LLM should avoid in its response \cite{DBLP:journals/corr/abs-2305-16807}. We apply this strategy by instructing the LLM to avoid common mistakes that can occur during the generation of POWL models, such as generating partial orders that violate irreflexivity. Moreover, we extend our few-shots demonstrations with common mistakes that should be avoided during the construction of each process. For example, a common mistake for the bicycle manufacturing process (cf. \autoref{lst:few-shots}) is to create a local choice between two activities ``reject order'' and ``accept order'' instead of modeling a choice between the complete paths that are taken in each case.
\end{itemize}

\subsection{Model Generation and Refinement}\label{sec:gen}
After receiving the LLM's response, the Python code snippet is extracted from the response, which might also include additional text (e.g., intermediate reasoning steps). If the code extraction is successful, then the extracted code is executed to generate the model. Executing code generated by an LLM involves multiple considerations to handle security risks and invalid results. The following strategies are implemented to ensure a safe environment for producing valid process models:
\begin{itemize}
    \item In order to eliminate the risk of executing unsafe code, we restrict the LLM to use the predefined functions we designed for the generation of POWL models. We employ a strict process to verify that the code strictly complies with the prompted coding guidelines, explicitly excluding the use of external libraries or constructs that may pose security threats. 
    \item Validation rules are in place to verify that the generated model conforms to the POWL specifications, such as the requirement that all partial orders respect transitivity and irreflexivity.
\end{itemize}

The framework supports displaying and exporting the generated POWL models as Petri nets or BPMN models for broader use within the business process management and process mining communities.

\paragraph{Refinement Loop} 
The framework also includes a refinement loop, allowing users to provide textual feedback on the generated models. Based on this feedback, the LLM is prompted to revise the model, ensuring continual improvement. 

\subsection{Error Handling}\label{sec:error}

Despite their advanced coding capabilities, LLMs are not immune to generating faulty code. We employ a robust error-handling mechanism tailored to mitigate potential inaccuracies and ensure the reliability of the generated process models. 

Recognizing the variability in the severity and implications of errors, we categorize them into two distinct groups:
\begin{itemize}
    \item \textit{Critical Errors:} These are severe issues that affect functionality or pose security risks, such as execution failures or major model validation violations. These errors must be resolved before the process can continue.
    \item \textit{Adjustable Errors:} These are less severe errors that affect the quality of the generated model, such as reusing sub-models within the same POWL model. These errors can be adjusted automatically, allowing for a degree of flexibility in their resolution. For example, the error of reusing submodels within the same POWL model can be automatically resolved by creating copies of the reused models. However, such intervention is approached with caution to prevent significant deviations from the behavior of the intended process.
\end{itemize}

Our framework uses an iterative error-handling loop, engaging the LLM to address the identified issues. A new prompt that details the error and requests the LLM to address it, along with the conversation history, are submitted to the LLM. This iterative cycle facilitates dynamic correction, leveraging the LLM's capabilities to refine and improve the generated code. 

Critical errors are handled by prompting the LLM repeatedly until a solution is found or the maximum allowed attempts are reached. If the LLM fails to fix the error after the allowed number of attempts, the system terminates the process and marks the model generation as unsuccessful. Adjustable errors are resolved automatically if the LLM fails to address them within a few iterations.

\subsection{Limitations}\label{sec:limit}
Our framework, while pioneering in leveraging LLMs for process modeling, has limitations. In this section, we outline areas for improvement and propose ideas for addressing them in future work.

\textit{Expanding Process Perspectives:} Our framework addresses the control-flow perspective of process modeling, omitting the data, resource, and operational perspectives, which are crucial for a comprehensive understanding of business processes. The inherent flexibility and understanding capabilities of LLMs present a significant potential for extending our framework to incorporate additional process perspectives.

\textit{Direct BPMN Generation:} The current implementation of our framework utilizes POWL for intermediate process representation. A possible direction for future research is the exploration of the direct generation of BPMN models without an intermediate process representation. This approach promises to offer greater flexibility in representing intricate process structures and dynamics and allows for the enrichment of process models with context-rich annotations. However, moving away from the structured guarantees provided by POWL necessitates the development of more advanced process model generation and validation techniques. 

\textit{Enhanced Interactivity:} We intend to enhance the model refinement loop to support more nuanced and interactive feedback mechanisms. For example, we aim to empower users to not only provide textual feedback on generated process models but also to manually edit the generated models.

\section{Tool Support}
\label{sec:tool}

In this section, we present the ProMoAI tool \cite{DBLP:conf/ijcai/KouraniB0A24} to support our process modeling framework. ProMoAI is available as a web application at \url{https://promoai.streamlit.app/}. 

Currently ProMoAI in integrated with three LLM providers: Google (\url{https://ai.google.dev/}), OpenAI (\url{https://openai.com/}), and DeepInfra (\url{https://deepinfra.com/}). Google offers the Gemini models, while OpenAI provides GPT and O1 models. DeepInfra supports popular open-source LLMs like Meta's LLaMa and Mistral, and it also enables custom model deployment.

\begin{figure*}[!t]
\centering
\includegraphics[width=0.75\textwidth]{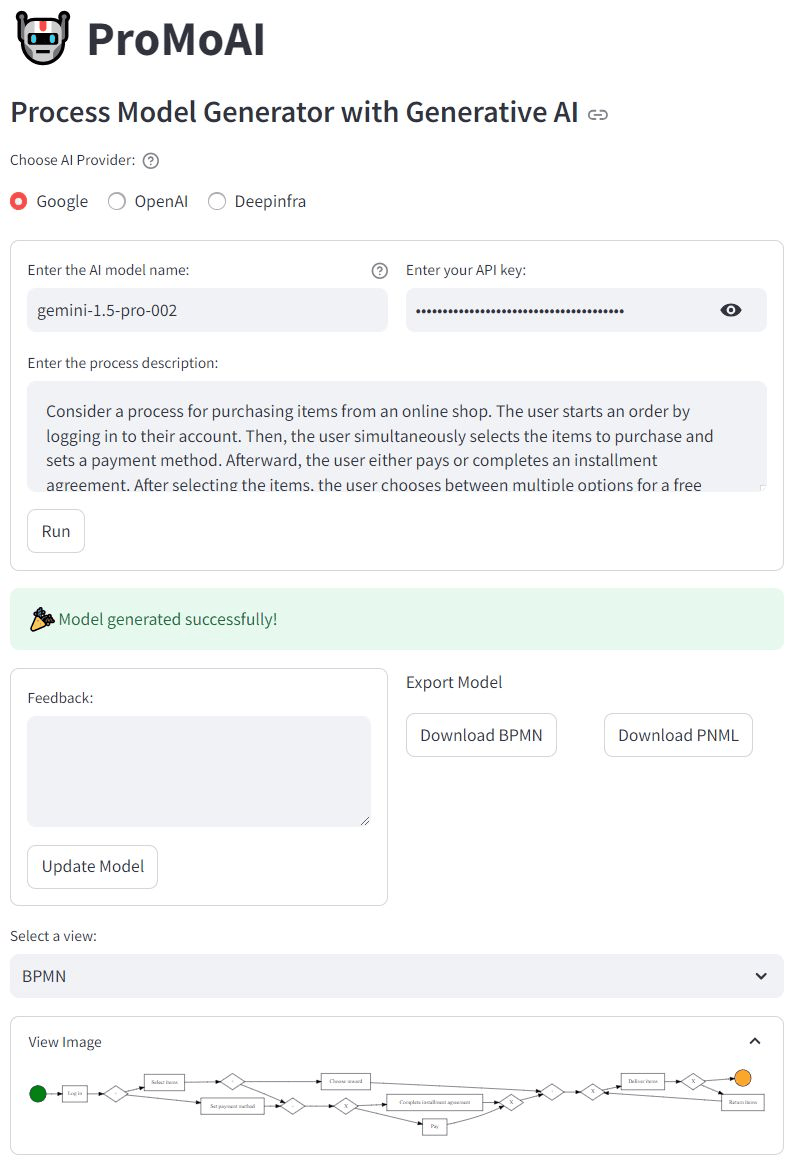}
\caption{A screenshot of ProMoAI.}
\label{fig:promoaiGeneralInterface}
\end{figure*}

A screenshot of ProMoAI is shown in \autoref{fig:promoaiGeneralInterface}. Initially, the application needs the specification of the LLM provider, the name of an LLM supported by the chosen provider, and an API key. The user needs to submit a textual description of the process to generate the initial process model. Once the process model is generated, it is presented to the user and can be viewed and downloaded as BPMN or Petri net. Following the initial process model generation, the user can provide feedback to iteratively improve the generated process model.

\section{Benchmarking State-of-the-Art Large Language Models}\label{sec:ev}

In this section, we present a comprehensive evaluation of our framework for process modeling using a diverse set of state-of-the-art LLMs. The objective is to assess the framework's capability to effectively generate high-quality business process models from natural language descriptions. Additionally, our evaluation serves as a benchmark for assessing the capabilities of state-of-the-art LLMs in a task that involves (1) modeling business processes starting from natural language descriptions, (2) generating executable code, (3) following instructions embedded in the input prompts, and (4) incorporating feedback to iteratively resolve errors and improve the quality of the outputs. 

We begin by detailing the experimental setup and design in \autoref{sec:ev:setup}, followed by an analysis of the results obtained in \autoref{sec:ev:results}. Note that all data and results are available at \url{https://github.com/humam-kourani/EvaluatingLLMsProcessModeling}.

\subsection{Experimental Setup and Design}\label{sec:ev:setup}
In this section, we detail the experimental setup and design. In \autoref{sec:ev:llmselect}, we discuss the selection of LLMs, while in \autoref{sec:ev:processdesign}, we outline the design of the processes we use in the evaluation. In \autoref{sec:ev:assessment}, we present our approach for assessing the quality of the generated process models. Finally, the configuration of framework's settings for error handling is described in \autoref{sec:ev:settings}.

\subsubsection{Selection of Large Language Models}\label{sec:ev:llmselect}
In order to achieve a comprehensive assessment of how different LLMs perform within our framework, we selected 16 LLMs that capture a wide spectrum of characteristics.

Our selection of LLMs includes models with different architectures (e.g., transformer-based, mixture-of-experts) and training methodologies (proprietary vs. open-source). We also considered size variation, ranging from smaller, faster models to very large models. Additionally, some models are optimized for specific tasks, such as code generation, deep reasoning, and following explicit prompt instructions. Finally, we ensure including models from different major AI vendors to ensure that our evaluation is comprehensive.

To thoroughly evaluate our framework, we selected the following 16 cutting-edge LLMs: 

\begin{itemize}
    \item \textbf{{\gptfour}}, \textbf{{\gptfouro}}, and \textbf{{\gptfouromini}}: Developed by OpenAI 
 (\url{https://openai.com/}), {\gptfour} is renowned for its advanced reasoning abilities, extensive knowledge base, and large context window. It excels in tasks requiring deep understanding and generation of coherent, contextually relevant text. {\gptfouro}, and {\gptfouromini} are optimized versions of {\gptfour}, offering faster performance and reduced computational requirements while maintaining high-quality outputs. 

    \item \textbf{{\oonepreview}} and \textbf{\oonemini}: The O1 series contains the latest advancements from OpenAI, outperforming previous models like {\gptfouro} across competitive benchmarks. \oonepreview\ is designed for deep reasoning, excelling in complex tasks like math, coding, and science. \oonemini\ is a smaller, faster, and more cost-effective version.

    \item \textbf{\geminipro} and \textbf{\geminiflash}: Developed by Google (\url{https://ai.google/}), the Gemini models are trained on diverse datasets. The Pro version is designed for enhanced reasoning capabilities, while the Flash variant emphasizes speed and efficiency.

    \item \textbf{\claude}: From Anthropic (\url{https://www.anthropic.com/}), {\claude} is designed for high performance with a focus on safety and reliability. It operates with a 200k token context window and is optimized for code generation and complex reasoning tasks.

    \item \textbf{\mistral}, \textbf{\codestral}, and \textbf{\openmixtral}: These models from Mistral AI (\url{https://mistral.ai/}) are designed for efficient training and inference with support for multilingual tasks. {\mistral} is top-tier reasoning model provided by Mistral AI for high-complexity tasks. {\openmixtral} is particularly noteworthy due to its mixture-of-experts architecture, which allows it to activate only a subset of its parameters during each inference, making it highly efficient. {\codestral} is optimized for generating and understanding code in a wide array of programming languages.

    \item \textbf{\llamathreeone} and \textbf{\llamathreetwo}: These open-source models from Meta (\url{https://ai.meta.com/}) are trained on extensive corpora and designed for instruction following and complex reasoning tasks. {\llamathreeone} is notable for its large parameter size, enhancing its capability to handle intricate tasks.

    \item \textbf{\nemotron}: Developed by Nvidia (\url{https://www.nvidia.com/}) as an advanced version of Meta's Llama-3.1-70B, Nemotron leverages Nvidia's cutting-edge hardware and fine-tuning techniques to offer high-performance capabilities. 

    \item \textbf{\qwen}: Developed by Alibaba Cloud (\url{https://www.alibabacloud.com/}), {\qwen} is a powerful open-source model known for its multilingual support and proficiency in handling complex instructions.

    \item \textbf{\wizard}: This is an advanced open-source model designed by Microsoft (\url{https://www.microsoft.com/}).

\end{itemize}

Note that for the open-source models \llamathreeone, \llamathreetwo, \nemotron, \qwen, and \wizard, we used the instances hosted by Deep Infra (\url{https://deepinfra.com/}).

\subsubsection{Design of Processes}\label{sec:ev:processdesign}

To evaluate the LLMs within our framework, we designed a set of 20 distinct pairs of process descriptions and their corresponding ground truth POWL model. We refer to these processes as p1, ..., p20 throughout this paper. Two processes were adapted from our previous work \cite{DBLP:conf/bpmds/KouraniB0A24}: an order handling process \cite{DBLP:conf/bpm/KouraniZ23} and a hotel service process \cite{DBLP:conf/bpm/BellanADGP22}. The remaining 18 processes were created to represent diverse business domains, including manufacturing, healthcare, finance, logistics, customer service, and more. This diversity ensures that the evaluation is not biased toward a specific industry or process type.

The processes were intentionally varied along several dimensions:

\begin{itemize}

    \item \textbf{Process Description Length}: Ranged from 79 to 230 words and from 525 to 1,567 characters.

    \item \textbf{Level of Detail}: Process descriptions varied in how explicitly they specified structural elements. While no descriptions were truly vague, certain structural details were deliberately left ambiguous to challenge the LLMs to reason about process structure. For example, some descriptions did not clarify whether tasks should be executed concurrently or sequentially, requiring the LLMs to infer the appropriate behavior based on the process context. 

    \item \textbf{Process Size}: The number of activities in the ground truth models ranged from 8 to 26, providing a mix of simple and complex processes. 

    \item \textbf{Structural Complexity}: The processes were designed to cover different levels of structural complexity for the three process constructs supported by POWL:

    \begin{itemize}
        \item \textit{Choices}: Included processes with skippable activities, simple choices between single activities, and choices involving complex sub-processes.
        \item \textit{Loops}: Incorporated processes with repeatable activities, simple loops over single activities, and loops involving complex sub-processes.
        \item \textit{Partial Orders}: Varied from highly sequential processes to those with high concurrency, as well as complex partial orders that contain non-hierarchical dependencies.
    \end{itemize}
\end{itemize}

\begin{figure*}[!t]
    \centering
    \begin{subfigure}[b]{0.45\textwidth}
        \centering
        \includegraphics[width=\textwidth]{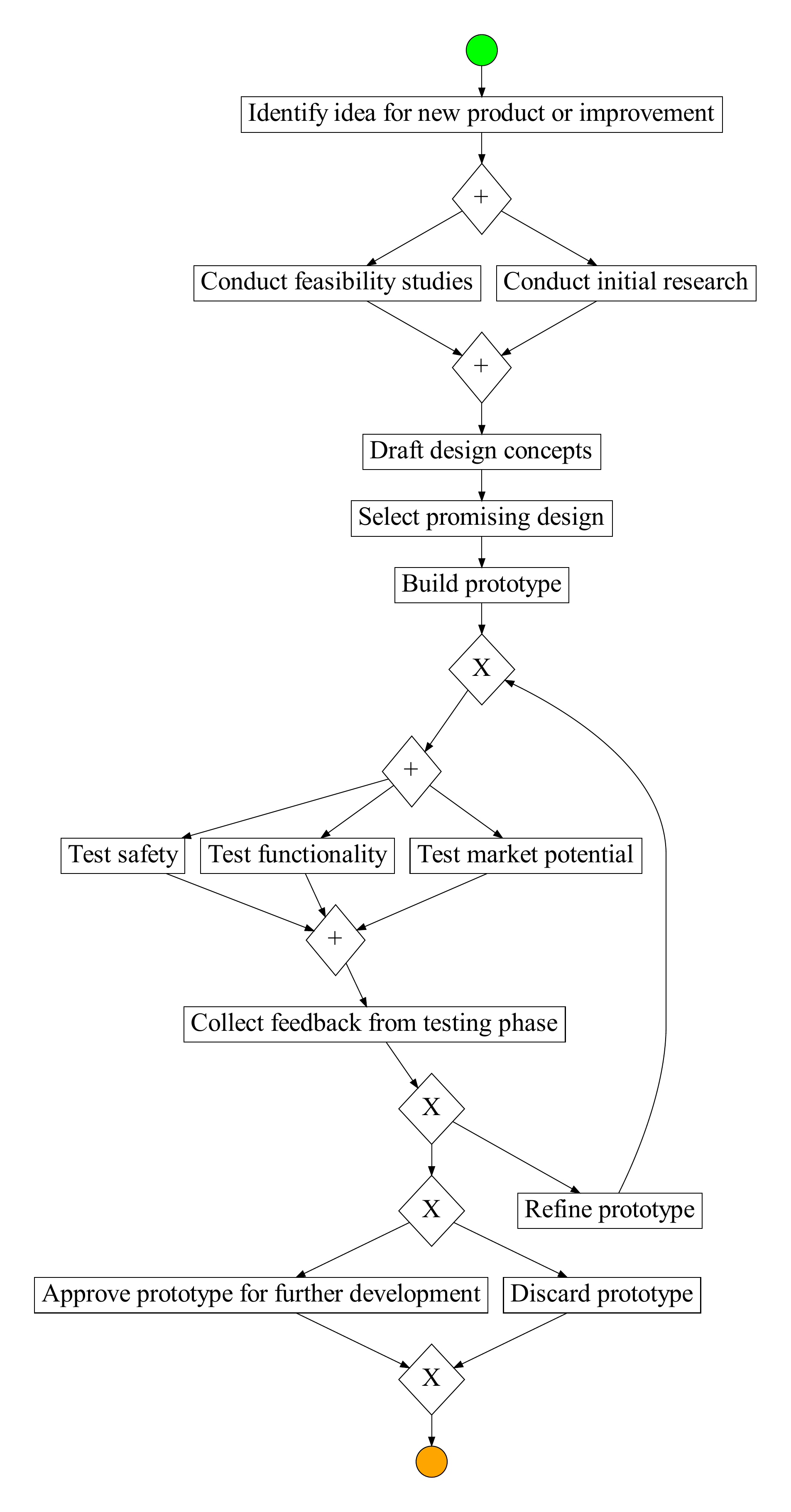}
        \caption{Ground truth (Score 0.98).}
        \label{fig:p9:gt}
    \end{subfigure}
    \hfill
    \begin{subfigure}[b]{0.50\textwidth}
        \centering
        \includegraphics[width=\textwidth]{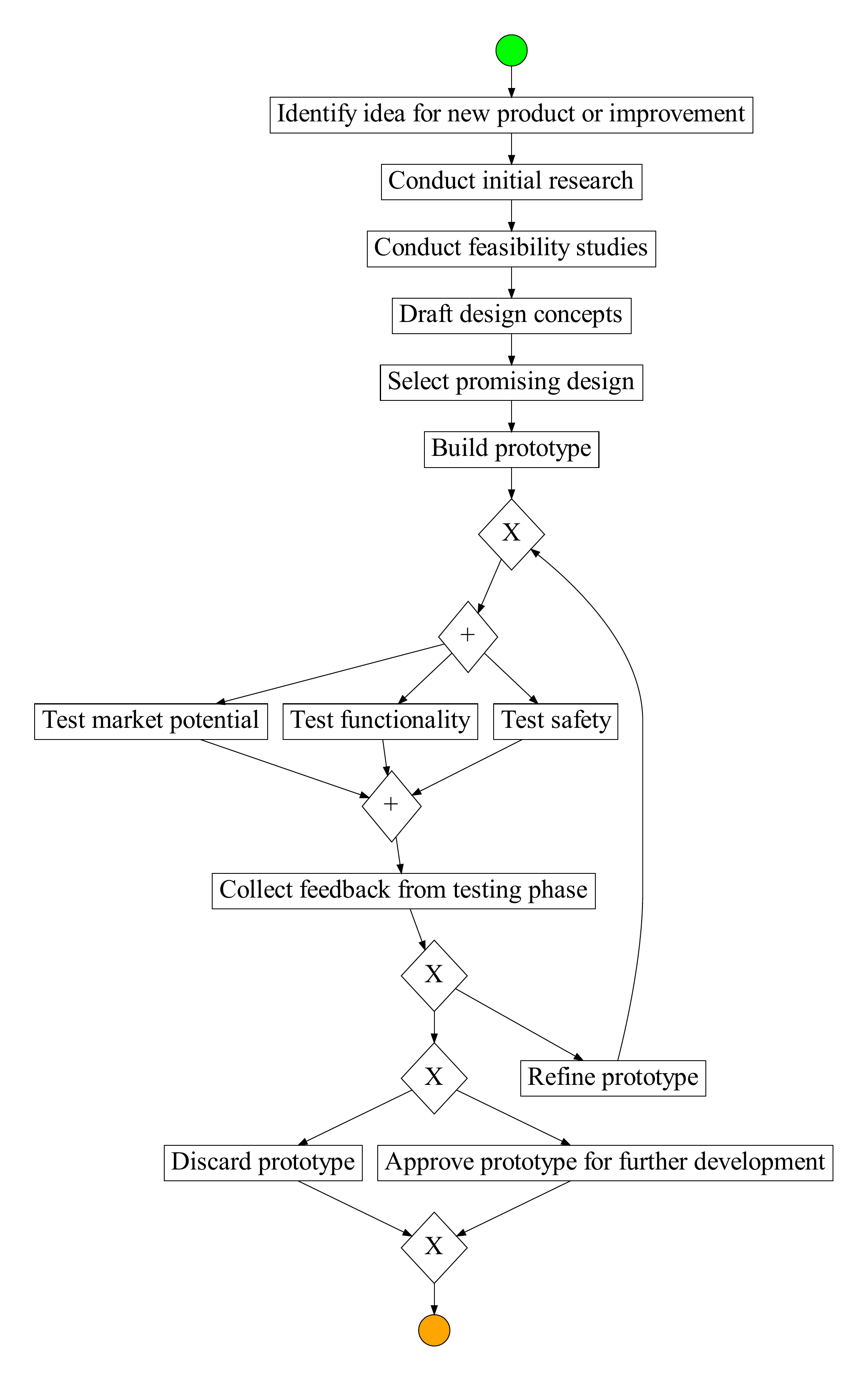}
        \caption{Generated with \oonemini\ (Score 0.97).}
        \label{fig:llm_generated_a}
    \end{subfigure}
    \caption{Ground truth and LLM-generated process models for process p9.}
    \label{fig:p9:1}
\end{figure*}

\begin{figure*}[!t]
    \centering
    \begin{subfigure}[b]{0.36\textwidth}
        \centering
        \includegraphics[width=\textwidth]{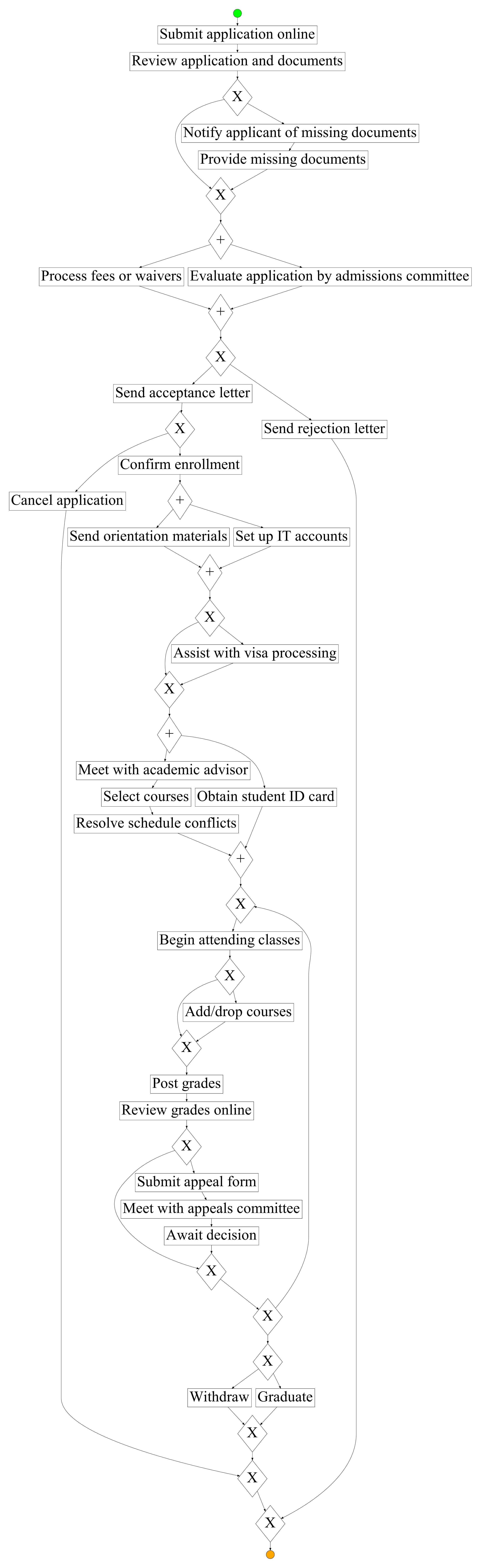}
        \caption{Ground truth (Score 0.99).}
        \label{fig:p18:gt}
    \end{subfigure}
    \hfill
    \begin{subfigure}[b]{0.36\textwidth}
        \centering
        \includegraphics[width=\textwidth]{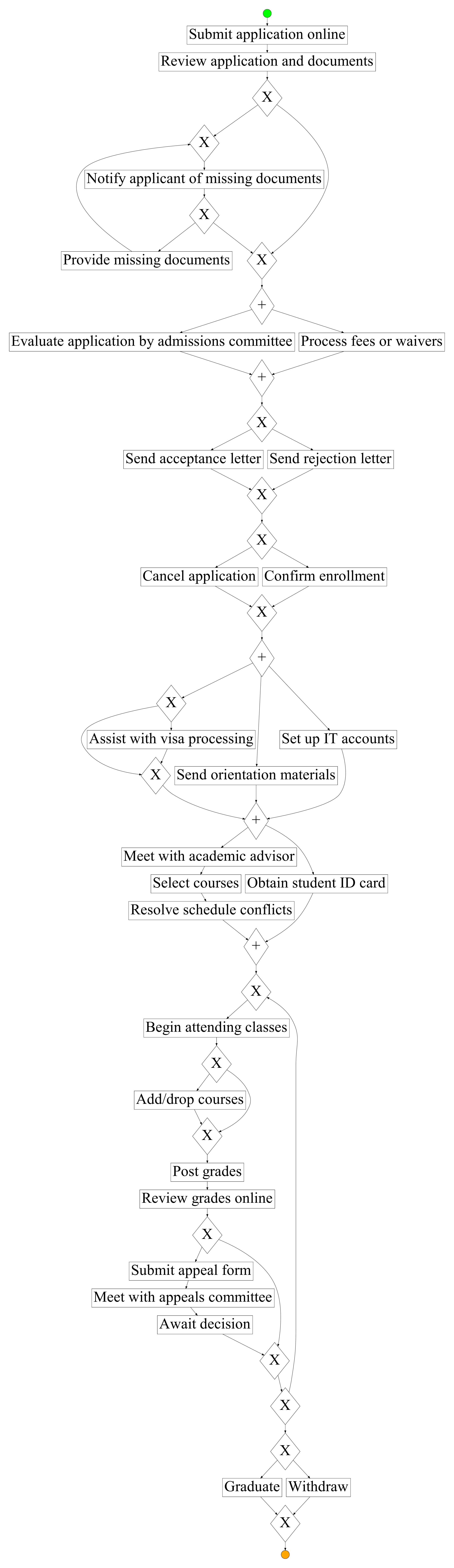}
        \caption{Generated with \claude\ (Score 0.93).}
    \end{subfigure}
    \caption{Ground truth and LLM-generated process models for process p18.}
    \label{fig:p18:1}
\end{figure*}

\autoref{lst:desc9} and \autoref{lst:desc18} show two example textual descriptions, illustrating variations in length and complexity. The corresponding ground truth models are shown in BPMN in \autoref{fig:p9:gt} and \autoref{fig:p18:gt}, respectively.

\begin{lstlisting}[caption={Textual description for process p9 (644 characters, 97 words).},label={lst:desc9}, frame=single, float, floatplacement='!t', basicstyle=\footnotesize\ttfamily]
The process starts with identifying an idea for a new product or improvement to an 
existing one. The R&D team conducts initial research and feasibility studies, 
followed by drafting design concepts. After selecting a promising design, a 
prototype is built using available materials and resources. The prototype 
undergoes various tests to assess its functionality, safety, and market potential. 
Feedback from the testing phase is collected, and the prototype may be refined 
accordingly. If a refinement is needed, then the testing phase is reinitiated. The 
process ends when the prototype is either approved for further development or 
discarded.
\end{lstlisting}

\begin{lstlisting}[caption={Textual description for process p18 (1567
characters, 230
words).},label={lst:desc18}, frame=single, float, floatplacement='!t', basicstyle=\footnotesize\ttfamily]
A university enrollment system involves the following steps:
Prospective students submit an application online.
The admissions office reviews the application and supporting documents.
If documents are missing, the applicant is notified to provide the missing items.
Upon receiving all documents, the application is evaluated by the admissions 
committee.
Concurrently, the finance department processes any application fees or waivers.
If the application is accepted, an acceptance letter is sent. Otherwise, a 
rejection letter is sent and the process ends.
After being accepted, the student must then confirm enrollment by a specified 
deadline; otherwise the application will be canceled.
If the student confirms, they receive orientation materials and the IT department 
sets up student accounts for email, online portals, and library access.
If the student is international, the international student office assists with visa 
processing.
The student obtains a student ID card and starts creating their study plan, which 
includes:
Meeting with an academic advisor.
Selecting courses.
Resolving any schedule conflicts.
The student begins attending classes.
Throughout each semester, the student may add or drop courses within the add/drop
period.
At the end of the semester, grades are posted, and the student can review them 
online.
If the student has any grievances, they can file an appeal, which includes:
Submitting an appeal form.
Meeting with the appeals committee.
Awaiting a decision.
The process repeats each semester until the student graduates or withdraws.
\end{lstlisting}

\begin{figure*}[!t]
    \centering
    \begin{subfigure}[b]{0.36\textwidth}
        \centering
        \includegraphics[width=\textwidth]{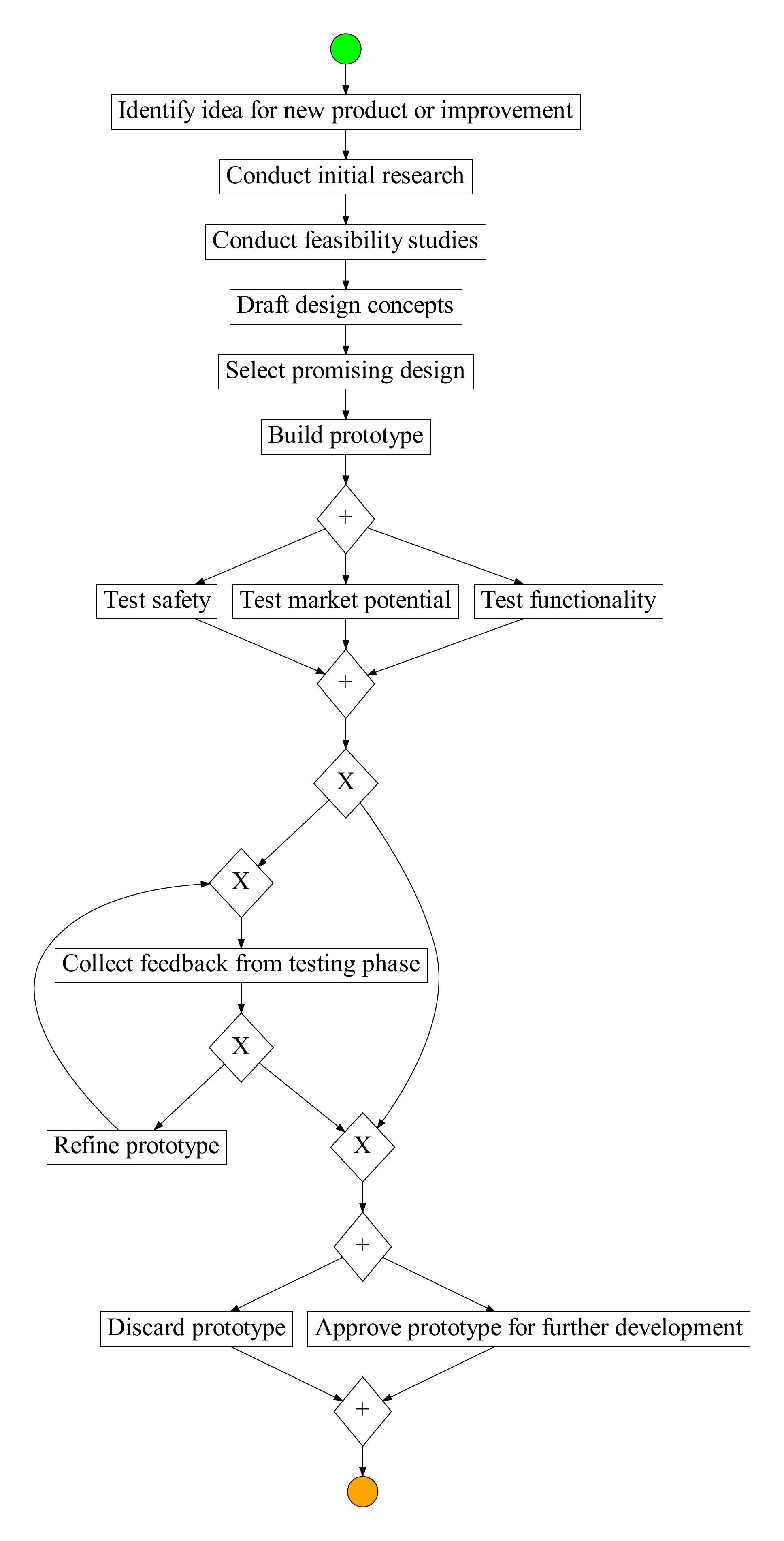}
        \caption{Generated with \llamathreetwo\ (Score 0.83).}
        \label{fig:ground_truth_b}
    \end{subfigure}
    \hfill
    \begin{subfigure}[b]{0.63\textwidth}
        \centering
        \includegraphics[width=\textwidth]{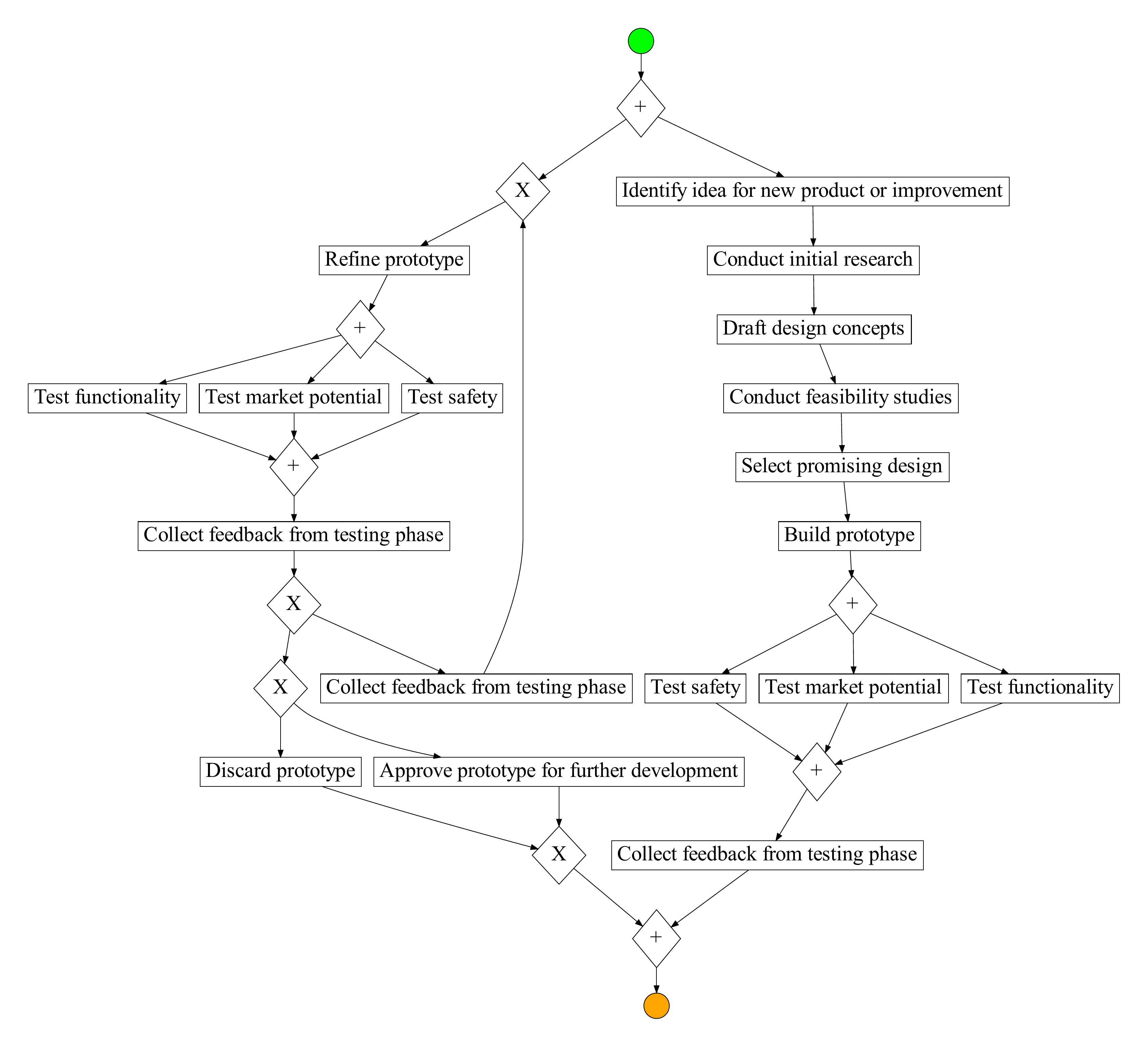}
        \caption{Generated with \codestral\ (Score 0.56).}
        \label{fig:llm_generated_b}
    \end{subfigure}
    \caption{LLM-generated process models for process p9.}
    \label{fig:p9:2}
\end{figure*}

\begin{figure*}[!t]
    \centering
    \begin{subfigure}[b]{0.43\textwidth}
        \centering
        \includegraphics[width=\textwidth]{Qwen2.5-72B-Instruct_p18.pdf}
        \caption{Generated with \qwen\ (Score 0.78).}
    \end{subfigure}
    \hfill
    \begin{subfigure}[b]{0.55\textwidth}
        \centering
        \includegraphics[width=\textwidth]{WizardLM-2-8x22B_p18.pdf}
        \caption{Generated with \wizard\ (Score 0.39).}
    \end{subfigure}
    \caption{LLM-generated process models for process p18.}
    \label{fig:p18:2}
\end{figure*}

\subsubsection{Assessment of Generated Models}\label{sec:ev:assessment}
To facilitate quantitative evaluation through conformance checking \cite{DBLP:conf/s-bpm-one/DunzerSMB19}, we simulated event logs from the ground truth POWL models. These logs serve as the basis for assessing the quality of the models generated by the LLMs using our framework.

\paragraph{Simulation of Event Logs}
The simulation was meticulously conducted to produce comprehensive and representative event logs from the ground truth models. This process was based on two key assumptions to ensure both feasibility and thorough coverage of possible process behaviors. First, we assume that all decision points within the process follow an equal distribution, meaning each possible choice at a decision point has an equal probability of being selected. This assumption simplifies the simulation by treating all paths through decision points as equally likely, thereby ensuring unbiased exploration of all possible \emph{trace variants} (i.e., distinct activity sequences). Second, we limit loops to a maximum length of two iterations, allowing only up to two executions of the loop's do-part. This constraint is necessary to prevent the generation of infinitely long event logs. By capping the loop iterations, we maintain a manageable size for the event logs while still capturing the essential repetitive behavior of the process. Under these two assumptions, each simulated event log includes one instance of every unique trace variant possible within the ground truth model, thereby ensuring comprehensive coverage of the process behavior.

\paragraph{Standardizing Activity Labels}
To ensure accurate conformance checking between LLM-generated process models and the ground truth event logs, it is essential to standardize activity labels. Without this standardization, inconsistencies in activity naming can result in misleading quality scores, as LLM-generated models may assign different labels to equivalent activities, introduce new activities, or combine  multiple activities into a single one. To address this, we extend the initial prompt with the list of activity labels of the ground truth log, presented in a random order to prevent the LLM from inferring sequential dependencies based on label ordering. We instruct the LLM to generate a process models using the same activity labels. While this may limit the LLM's flexibility in naturally expressing process steps, it mirrors real-world scenarios where key process steps are known, even if the overall process model is not formally defined. Moreover, this standardizing is crucial for enabling conformance checking and automated quantitative assessment.

\paragraph{Quality Score Computation}
We assess the quality of the LLM-generated models in relation to the ground truth event logs by calculating fitness \cite{fitness} and precision \cite{precision} metrics using the PM4Py library \cite{DBLP:journals/simpa/BertiZS23}. The overall quality score for each generated model is determined as the harmonic mean of the fitness and precision values. High scores (approaching 1.0) indicate a high level of conformance with the ground truth event logs, whereas lower scores (closer to 0.0) reflect poor model quality.

\subsubsection{Error Handling Settings}\label{sec:ev:settings}
Our framework incorporates an error-handling loop as described in \autoref{sec:error}. We use a threshold of ten iterations for addressing adjustable errors. If errors persist beyond these ten iterations, the framework automatically attempts to fix them, which may impact the overall quality of the generated models. Afterwards, if errors remain, an additional five iterations are permitted. By setting high iteration thresholds, we aim to assess the LLMs' proficiency in understanding and correcting their outputs based on feedback and to ensure fairness by providing ample opportunity for each model to succeed.

\subsection{Results and Analysis}\label{sec:ev:results}
In this section, we present the results of our evaluation, focusing on the error-handling performance (\autoref{sec:ev:result:error}), quality of the generated models (\autoref{sec:ev:result:quality}), time efficiency (\autoref{sec:ev:result:time}), and overall observations (\autoref{sec:ev:result:summary}). 

Example models generated for the processes $p9$ and $p18$ are shown in \autoref{fig:p9:1}, \autoref{fig:p9:2}, and \autoref{fig:p18:1}, \autoref{fig:p18:2}.

\subsubsection{Error Handling Performance}\label{sec:ev:result:error}
\begin{table*}[!t]
    \centering
    \caption{Error handling performance metrics.}
    \label{tab:error_handling_performance}
    \resizebox{\textwidth}{!}{   
        \begin{tabular}{lcccc}
            \toprule
            \multirow{2}{*}{\textbf{Model}} & \textbf{Avg. Num.} & \textbf{Num. Cases} & \textbf{Num. Cases with} & \textbf{Num. Cases} \\
             & \textbf{Iterations} & \textbf{without Errors} & \textbf{Auto-Adjustment} & \textbf{with Failures}\\
            \midrule
            \claude & 1.35 & 16 & 0 & 0 \\
            \oonemini & 1.4 & 14 & 0 & 0 \\
            \oonepreview & 1.5 & 14 & 0 & 0 \\
            \geminipro & 1.95 & 13 & 0 & 0 \\
            \gptfouro & 2.25 & 9 & 0 & 0 \\
            \llamathreeone & 2.55 & 9 & 0 & 0 \\
            \mistral & 2.6 & 10 & 0 & 0 \\
            \llamathreetwo & 2.95 & 8 & 0 & 0 \\
            \geminiflash & 3.3 & 4 & 0 & 0 \\
            \openmixtral & 3.6 & 10 & 3 & 1 \\
            \gptfour & 3.9 & 2 & 0 & 0 \\
            \codestral & 3.9 & 7 & 2 & 1 \\
            \gptfouromini & 4.05 & 6 & 3 & 0 \\
            \nemotron & 4.05 & 3 & 1 & 0 \\
            \qwen & 4.65 & 4 & 2 & 0 \\
            \wizard & 5.2 & 8 & 5 & 0 \\
            \bottomrule
        \end{tabular}
    }
\end{table*}

We analyzed how each LLM performed in terms of the number of iterations required to generate a valid process model without errors. In \autoref{tab:error_handling_performance}, we report the average number of iterations required to produce a valid model by each LLM, the number of cases where only one iteration was needed (indicating no errors), the number of instances where more than ten iterations were necessary (indicating failure to resolve adjustable errors), and the number of cases where no valid model was generated after 15 iterations (indicating failure to resolve critical errors).

\claude\ demonstrated exceptional performance, with an average of 1.35 iterations and 16 out of 20 instances where only one iteration was sufficient to produce a valid model. This performance was closely mirrored by the O1 models and \geminipro, which averaged less than two iterations and managed to produce models without any errors in 14 and 13 instances, respectively. These results suggest that these models possess robust reasoning capabilities and an ability to generate high-quality results on the first attempt.

Among the models evaluated, \wizard\ stood out with the highest average number of iterations (5.2) and five instances where more than 10 iterations were needed. However, it is notable that despite these higher iteration counts, \wizard\ did not experience any failures. Models such as \gptfour\ and \geminiflash\ demonstrated low single-iteration success rates (2 and 4 cases, respectively); however, these models were able to achieve a moderate average number of iterations and managed to resolve both adjustable and critical errors. This underscores their effectiveness in incorporating feedback to iteratively improve the quality of their outputs. In contrast, other models showed a higher tendency for requiring additional iterations and occasional failures. \openmixtral\ necessitated manual adjustments in three instances and failed to generate a valid model once after 15 iterations. Similarly, \codestral\ encountered two such instances of manual adjustments and one failure. The results for \codestral, which is optimized for code generation, were particularly surprising. Despite its specialization, this model performed worse than the other two Mistral models we considered. This suggests that task-optimized models face challenges on their primary task when applied to new domains.

\subsubsection{Quality of the Generated Models}\label{sec:ev:result:quality}
As described in \autoref{sec:ev:assessment}, we assess the quality of the generated process models using the harmonic mean of fitness and precision scores obtained from conformance checking the models against the ground truth event logs. The average obtained quality scores for each LLM are reported in \autoref{tab:ev:quality}.

\begin{table}[!t]
    \centering
    \caption{Average quality scores.}
    \label{tab:ev:quality}
    \begin{tabular}{lc}
        \toprule
        \textbf{Model} & \textbf{Avg. Score} \\
        \midrule
        Ground Truth & 0.98 \\
        \midrule
        \claude & 0.93 \\
        \oonepreview & 0.92 \\
        \oonemini & 0.91 \\
        \geminipro & 0.87 \\
        \llamathreeone & 0.86 \\
        \nemotron & 0.83 \\
        \llamathreetwo & 0.80 \\
        \qwen & 0.80 \\
        \mistral & 0.78 \\
        \gptfour & 0.76 \\
        \gptfouro & 0.76 \\
        \gptfouromini & 0.74 \\
        \wizard & 0.73 \\
        \codestral & 0.73 \\
        \geminiflash & 0.73 \\
        \openmixtral & 0.72 \\
        \bottomrule
    \end{tabular}
\end{table}

The results demonstrate significant variation in the quality of generated models across different LLMs. \claude\ leads the pack with an impressive average score of 0.93, closely approaching the average score of the ground truth models (0.98)\footnote{A perfect score (1.0) is unattainable in many cases due to the presence of loops within the process models, which inherently allow for infinite behavior.}. Following \claude, the \oonepreview\ and \oonemini\ models also exhibit strong performance, achieving average scores of 0.92 and 0.91, respectively. These high scores underscore the effectiveness of the O1 series in generating accurate and reliable results, attributable to their advanced reasoning capabilities.

A closer examination of the results reveals the following observations and patterns:

\textbf{(1) Positive Correlation Between Iteration Performance and Quality:} 
There is a notable correlation between error-handling performance and model quality. Models like \claude, \oonemini, \oonepreview, and \geminipro, which produced less errors and required fewer iterations to generate valid process models, also achieved higher quality scores. Conversely, models that required more iterations or experienced failures, such as \openmixtral, \codestral, and \wizard, tended to have lower scores, reflecting potential compromises in model accuracy due to extended correction processes. This indicates that the ability to generate error-free outputs promptly contributes to the overall quality of the process models. 

\textbf{(2) Quality Consistency on Average:}
Despite the inherent non-determinism in LLM outputs, we observe consistent quality trends within similar groups of models. Specifically, the O1 models achieved average scores of 0.92 and 0.91, while the two instances of the Llama 3.1 family scored on average 0.86 and 0.83. The three models within the GPT-4 family recorded average scores ranging from 0.74 to 0.76. Additionally, across four independent runs of \geminipro\ (cf. \autoref{sec:selfev}), the model consistently exhibited similar performance with average scores ranging between 0.86 and 0.88. The observed patterns suggest that, although re-executing the same or similar LLMs on identical tasks can yield varying results, the average quality remains stable within each model group. 

\textbf{(3) Impact of Speed Optimization:}
Models optimized for speed exhibited varied behaviors in terms of quality scores. Notably, the OpenAI models \oonemini\ and \gptfouromini\ maintained quality scores comparable to their base counterparts, \oonepreview\ and \gptfouro, respectively. This consistency suggests that optimization for speed in these models does not significantly compromise the quality of their outputs. In contrast, \geminiflash\ demonstrated lower quality scores compared to the \geminipro\ variant. However, Google promotes \geminiflash\ as a small, lightweight model designed for tasks where speed and cost-effectiveness matter the most; therefore, it is unfair to compare its performance with the Pro model. These examples illustrate that speed optimization strategies can significantly differ based on the desired balance between speed enhancement and quality preservation, either aiming for moderate speed improvements with minimal quality loss or prioritizing speed as the primary optimization goal.

\subsubsection{Time Efficiency}\label{sec:ev:result:time}

Time efficiency is a critical factor in evaluating the practicality and scalability of different LLMs in our framework. We assessed the performance of each LLM by measuring both the average cumulative time taken across all iterations to generate the process model and the average time per a single iteration. \autoref{tab:ev:time_efficiency} reports these metrics, ordered by the average total time in ascending order.

\begin{table*}[!t]
    \centering
    \caption{Time efficiency metrics.}
    \label{tab:ev:time_efficiency}
    \resizebox{\textwidth}{!}{  
        \begin{tabular}{lccc}
            \toprule
            \textbf{Model} & \textbf{Avg. Total Time (sec)} & \textbf{Avg. Time per Iteration (sec)}\\
            \midrule
            \geminiflash & 14.51 & 4.03\\
            \claude & 23.63 & 16.88 \\
            \geminipro & 24.86 & 12.06 \\
            \codestral & 38.27 & 7.98 \\
            \openmixtral & 52.88 & 12.20 \\
            \oonemini & 55.29 & 39.16 \\
            \gptfouromini & 56.20 & 12.11 \\
            \llamathreeone & 67.86 & 24.69 & \\
            \llamathreetwo & 72.97 & 21.56 & \\
            \gptfouro & 78.98 & 33.72 \\
            \gptfour & 108.55 & 26.56 \\
            \qwen & 126.98 & 20.59 \\
            \oonepreview & 145.48 & 90.84 \\
            \nemotron & 167.98 & 38.20 \\
            \mistral & 169.66 & 59.46 \\
            \wizard & 181.47 & 29.57 \\
            \bottomrule
        \end{tabular}
    }
\end{table*}
\geminiflash\ and \codestral\ emerge as the fastest models in terms of per-iteration processing speed, with average times of 4.03 seconds and 7.98 seconds respectively. However, their rapid response times come at the cost of requiring more iterations and a compromise in quality, diminishing their overall effectiveness. In contrast, \claude\ and \geminipro\ exemplify models that strike an excellent balance between speed and quality. \claude\ is particularly noteworthy, ranking second in total average time at 23.63 seconds while ranking first in average quality at a score of 0.93. Similarly, \geminipro\ demonstrate robust performance with average total times around 24.86 seconds and a strong average quality score of 0.87. These models achieve their efficiency not only through relatively fast per-iteration times but also by requiring less iterations and effectively handling errors.

The reasoning models of the O1 series (\oonemini\ and \oonepreview) exhibit longer per-iteration processing times (39.16 seconds and 90.84 seconds respectively). Despite their slower speeds per iteration, they maintain moderate total processing times due to their high efficiency in producing valid models quickly, minimizing the need for error handling iterations. \oonemini, in particular, achieves a similar quality score to \oonepreview\ but at a substantially lower total time, making it a highly efficient option within the O1 series. This pattern extends to other OpenAI models, where \gptfouromini\ outperforms \gptfour\ and \gptfouro\ in speed while maintaining similar quality levels.

At the lower end of the efficiency spectrum, \wizard, \mistral, and \nemotron\ stand out as the least time-efficient models, with average total times between 181.47 seconds and 167.98 seconds.

\subsubsection{Summary of Results}\label{sec:ev:result:summary}
In summary, our evaluation demonstrates that \claude\ stands out as the best-performing LLM in our framework, delivering the highest quality process models with minimal iterations and efficient error handling. \geminipro\ follows closely behind, with lower quality scores but also offering a good balance between quality and efficiency. The O1 series models (\oonemini\ and \oonepreview) exhibit excellent performance but fall behind in the overall time performance. Speed-optimized models like \geminiflash\ provide faster per-iteration times but often require more iterations and may compromise on quality. 

These findings highlight the trade-offs between speed, iteration count, quality, and cost across different LLMs, emphasizing the importance of selecting the right LLM to achieve the best balance. Note that we did not include cost information in our evaluation due to the rapidly changing landscape of LLM pricing. However, when taking cost into account, \geminipro\ offered the best trade-off as it was freely available through APIs for limited, small-scale testing.

\section{Evaluating LLM Self-Improvement Strategies}\label{sec:ev2}

In this section, we investigate the ability of LLMs to improve the quality of their outputs through self-improvement strategies. We aim to examine how LLMs can autonomously evaluate, refine, and optimize their performance within our framework for process modeling. Specifically, we explore three techniques: \emph{self-evaluation} (cf. \autoref{sec:selfev}), \emph{self-optimization of input} (cf. \autoref{sec:selfImproveInput}), and \emph{self-optimization of output} (cf. \autoref{sec:selfImproveOutput}). 

We primarily use \geminipro\ in this section due to its optimal trade-off between performance, time, and cost, as discussed in \autoref{sec:ev}. We utilize the same set of 20 processes previously described in \autoref{sec:ev}. All results are available at \url{https://github.com/humam-kourani/EvaluatingLLMsProcessModeling}.

\subsection{LLM Self-Evaluation}\label{sec:selfev} 

In this section, we explore the potential of self-evaluation by LLMs to enhance the quality of their outputs within our process modeling framework. 

Self-evaluation capitalizes on the reasoning capabilities of LLMs, enabling them to assess and potentially refine their own outputs. The aim is to reduce errors and hallucinations, thereby increasing the reliability and accuracy of LLM outputs \cite{DBLP:conf/acl/ZhangPTZJSMM24}. Given the inherent non-determinism of LLM outputs, where responses can vary significantly across different sessions \cite{DBLP:journals/corr/abs-2407-10457}, self-evaluation might help in stabilizing outputs and mitigating variability.

Self-evaluation can be utilized in several ways: providing a quality score to the user to indicate the LLM’s confidence in its own answers, generating multiple candidate outputs and selecting the best one, or combining several responses into a single, more robust output. In our framework, we implement LLM self-evaluation by generating multiple candidate process models for each process description, and then we instruct the LLM to assess them based on predefined criteria and select the best model.

\subsubsection{Implementation and Experimental Setup}
For each process description, we conducted four independent runs, generating four candidate models (labeled \texttt{R1} to \texttt{R4}). After generating the four candidate models, we tasked the LLM with self-evaluating them and selecting the best one. In addition to \geminipro, we replicated the experiment using \geminiflash\ to assess whether the impact of self-evaluation differs between high-performing and lower-performing LLMs.

We crafted a comprehensive prompt that includes the initial task description, the specific process description, the four generated candidate models, and a request for the LLM to self-evaluate these candidates and provide an overall quality score for each. Additionally, we computed the ground truth quality scores as described in \autoref{sec:ev:assessment} to validate whether the LLM's selections align with these scores, thereby verifying the effectiveness of its self-evaluation capabilities.

\paragraph{Evaluation Criteria} 
We instructed the LLM to evaluate the generated models based on sets of criteria to ensure a comprehensive assessment of each candidate model's quality according to different perspectives. We created two sets of evaluation criteria: a general set (cf. \autoref{lst:general_evaluation}) and a conformance-based set (cf. \autoref{lst:conformance_evaluation}). The first set of criteria focuses on broader aspects of model quality, while the second set aligns with the quality metrics used to compute the ground truth scores.

\begin{lstlisting}[caption={Evaluation criteria in the prompt for general self-evaluation.},label={lst:general_evaluation},frame=single, float, floatplacement='!t', basicstyle=\footnotesize\ttfamily]
**Evaluation Criteria:**
- **Behavior Accuracy:** How accurately does the model capture the intended process 
behavior?
- **Completeness:** Does the model include all necessary activities as described?
- **Correctness:** Are the control flows (e.g., partial orders, choices, loops) 
correctly implemented?
\end{lstlisting}
    
\begin{lstlisting}[caption={Evaluation criteria in the prompt for conformance-based self-evaluation.},label={lst:conformance_evaluation}, frame=single, float, floatplacement='!t', basicstyle=\footnotesize\ttfamily]
**Evaluation Criteria:**
- **Fitness:** Evaluate how well the process model can reproduce the behaviors of 
the process according to the process description.
- **Precision:** Evaluate the extent to which the process model exclusively 
represents behaviors that are allowed in the process according to the process 
description.
\end{lstlisting}

\subsubsection{Results and Discussion}

\begin{table*}[!t]
\centering
\caption{Impact of LLM self-evaluation on process model quality. Under both sets of evaluation criteria (general and conformance-based), we report the average quality scores of the process models before self-evaluation and selection, the number of instances where the LLM's selected process models are a subset of the best process models based on the quality assessment, the number of instances of exact matches, and the average quality scores of the process models post LLM self-evaluation and selection.}
\label{tab:self_evaluation_results}
\resizebox{\textwidth}{!}{
\begin{tabular}{lccccc}
\toprule
\multirow{2}{*}{\textbf{LLM}} & \textbf{Avg. Quality Without} & \textbf{Evaluation} & \textbf{Subset} & \textbf{Exact} & \textbf{Avg. Quality} \\
& \textbf{Self-Eval. (R1-R4)} & \textbf{Criteria} & \textbf{Match} & \textbf{Match} & \textbf{With Self-Eval.} \\
\midrule
\multirow{2}{*}{\geminipro} & \multirow{2}{*}{0.86-0.88} & General & 15/20 & 5/20 & 0.91 \\
 & & Conformance & 15/20 & 7/20 & 0.91 \\
 \midrule
\multirow{2}{*}{\geminiflash} & \multirow{2}{*}{0.73-0.75} & General & 4/20 & 0/20 & 0.72 \\
& & Conformance & 3/20 & 0/20 & 0.72 \\
\bottomrule
\end{tabular}
}
\end{table*}

The results of the LLM self-evaluation experiments are summarized in \autoref{tab:self_evaluation_results}. It reports the average quality scores of the process models without self-evaluation, the number of instances where the LLM's selected process models are a subset of the best process models based on the quality assessment, the number of instances of exact matches, and the average quality scores of the process models post LLM self-evaluation and selection. To account for minor variations in quality scores, we applied a 0.02 buffer when determining the best process models in our match calculations.


For \geminipro, the initial average quality of the candidate process models ranged between 0.86 and 0.88. After applying self-evaluation and selection, the average quality increased to 0.91, indicating an overall improvement. However, these improvements were not consistent across all cases. In 5 out of 20 cases, the LLM's selected best models did not align with the selection based on the quality assessment. Despite these discrepancies, the increase in average quality suggests that the self-evaluation strategy was generally effective for \geminipro. The LLM's ability to select models that improved the overall average quality demonstrates its capacity to critically assess its outputs.

In contrast, \geminiflash\ exhibited lower performance. The initial average quality of its candidate models was between 0.73 and 0.75. After self-evaluation, the average quality slightly decreased to 0.72. The LLM selected the wrong models in 16 or 17 out of 20 cases, depending on the evaluation criteria used. Although the final impact on average quality was not high, these results suggest that the self-evaluation strategy may be disadvantageous for \geminiflash. 

Regarding the evaluation criteria, the results indicate that there was no significant difference between using the general evaluation criteria and the conformance-based criteria for both LLMs. The LLMs made similar assessments regardless of the prompted criteria, suggesting that the choice of evaluation criteria may not significantly influence the effectiveness of the self-evaluation.


In summary, the effectiveness of employing LLM self-evaluation to select the best output among multiple candidates appears to be highly dependent on the selection of the LLM. Our results highlight that this strategy might be more beneficial for higher-performing models like \geminipro; however, even for such models, the improvements are not consistent across all cases. Consequently, while LLM self-evaluation holds potential, the question of whether the potential improvements justify the additional time and costs remains unresolved, as the performance gains vary depending on the chosen LLM.

We acknowledge that these findings are limited as we only used two LLMs in our experiments. It is also important to note that our conclusions cannot be generalized to broader applications. Given that all candidate process models were generated by the same LLM, their quality levels are inherently close. We anticipate that LLM self-evaluation might yield better results when applied to outputs with larger disparities in quality.

\subsection{LLM Self-Optimization of Input}\label{sec:selfImproveInput}

In this section, we investigate the potential of LLMs to enhance the quality of process models by self-optimizing the input process descriptions. The hypothesis is that by allowing LLMs to refine and enrich the initial process descriptions, they might produce higher-quality process models. 

\subsubsection{Implementation and Experimental Setup}
For each process description, we created two additional versions: a summarized version that retains 50-80\% of the original description’s length (medium-length version), and a compact, very high-level version with 15-35\% of the original length (short version). By crafting shorter versions of the process descriptions, we aimed to introduce varying levels of detail and specificity. The motivation behind this was to give the LLM more latitude to enhance and clarify the descriptions, potentially leading to improved process models upon self-optimization.

\begin{lstlisting}[caption={Prompt for input optimization.},label={lst:input_optimization}, frame=single, float, floatplacement='!t', basicstyle=\footnotesize\ttfamily]
You are provided with a process description. Your task is to optimize this
description to make it richer and more detailed, while ensuring that all additions
are relevant, accurate, and directly related to the original process. The goal is
to make the description more comprehensive and suitable for process modeling
purposes.

Possible areas for enhancement include:
- **Detail Enhancement:** Add specific details that are missing but crucial for
understanding the process flow.
- **Clarity Improvement:** Clarify any ambiguous or vague statements to ensure that
the description is clear and understandable.
- **Explicit Process Constructs:** Rephrase parts of the description to explicitly
incorporate constructs. For example, change `X happens in most cases' to `there
is an exclusive choice between performing X or skipping it'.
\end{lstlisting}

For each of the three versions (long, medium-length, and short), we instructed \geminipro\ to improve the description using the prompt illustrated in \autoref{lst:input_optimization}. This prompt encourages the LLM to focus on enriching the description by adding relevant details, clarifying ambiguities, and making process constructs explicit, without introducing unrelated information.

\subsubsection{Results and Discussion}

We evaluated the quality of the process models generated from both the original and the LLM-improved process descriptions. A summary of the results is presented in \autoref{tab:ev:improveinput}.

\begin{table*}[!t] 
\centering 
\caption{Comparison of average quality scores before and after LLM self-improvement of process descriptions.} \label{tab:ev:improveinput} 
\resizebox{\textwidth}{!}{
\begin{tabular}{lccc} 
\toprule 
\multirow{2}{*}{\textbf{Description Length}} & \textbf{Avg. Quality Before} & \textbf{Avg. Quality After} & \textbf{Cases With} \\
  & \textbf{Self-Improvement} & \textbf{Self-Improvement} & \textbf{Increased Quality} \\
\midrule
Long (Original) & 0.87 & 0.79 & 6/20 \\ 
Medium-Length (50-80\%) & 0.75 & 0.82 & 11/20 \\
Short (15-35\%) & 0.78 & 0.72 & 8/20\\
\bottomrule 
\end{tabular}}
\end{table*}

Contrary to our hypothesis, our investigation reveals that, for our specific application and framework, LLM self-optimization of input does not yield consistent benefits and may even be counterproductive. For the original long descriptions, the average quality score was 0.87, which decreased to 0.79 after LLM self-optimization, with only 6 out of 20 cases showing improvement. For the medium-length descriptions, the average score increased from 0.75 to 0.82 post-optimization, with 11 cases showing improvement and 9 not. For the short descriptions, the average score decreased from 0.78 to 0.72 after self-optimization, with improvements in only 8 cases.

These results suggest that the changes introduced by the LLM during the self-optimization process did not systematically contribute to better model quality. While LLMs can generate coherent text, they may lack the specific domain knowledge required to accurately enrich process descriptions in a way that leads to better process models. 

In conclusion, our findings suggest that relying on LLMs to autonomously enhance process descriptions without domain-specific guidance or constraints does not effectively improve the resultant process models.

\subsection{LLM Self-Optimization of Output}\label{sec:selfImproveOutput}

In this section, we examine the potential of LLMs to enhance the quality of their outputs by self-optimizing the generated process models. The underlying hypothesis is that by enabling LLMs to critically evaluate and refine their own outputs, they may identify and correct flaws, leading to higher-quality process models.

\subsubsection{Implementation and Experimental Setup}

To investigate this approach, we extended our previous experiments by instructing the LLMs to perform self-optimization on their initial outputs. Specifically, after generating the initial process model from the process description, we prompted the LLM to critically evaluate the model against the initial description and improve it accordingly. The prompt used is shown in \autoref{lst:output_optimization}. To prevent unnecessary alterations that might degrade the model's quality, we crafted the prompt with intentional restrictiveness, emphasizing that the LLM should only make genuinely beneficial changes and encouraging it to retain the same model if no areas for improvement are identified. This approach addresses the tendency of LLMs to respond affirmatively to requests, even when they may lack the knowledge or capability to perform the task, potentially leading to unintended \emph{hallucinations}.

\begin{lstlisting}[caption={Prompt for output optimization.},label={lst:output_optimization}, frame=single, float, floatplacement='!t', basicstyle=\footnotesize\ttfamily]
Could you further improve the model? Please critically evaluate the process model 
against the initial process description and improve it accordingly **only where 
genuinely beneficial**. If you see no significant areas for enhancement, it is 
perfectly acceptable to return the same model without any changes. 
\end{lstlisting}

We conducted experiments using three different LLMs: \geminipro, \geminiflash, and \gptfouro. The selection of these models was motivated by the desire to assess whether self-optimization could yield more significant improvements in less-performing models (\geminiflash\ and \gptfouro) compared to a high-performing model (\geminipro).

For each process in our dataset, we generated the initial process model using each LLM and then applied the self-optimization prompt. We then evaluated the quality of the initial and improved models. For this experiment, we disabled the error refinement loop. Instead, in cases where errors occurred, we repeated the same self-optimization prompt multiple times until an error-free response was achieved. By bypassing error-handling loops, we aimed to avoid distracting the LLM with error correction, which might steer it away from the primary goal of enhancing the already successfully generated model.

\subsubsection{Results and Discussion}

The results of the experiment are summarized in \autoref{tab:res:self-imp;output}. This table includes, for each LLM, the average quality scores before and after LLM self-optimization of output, as well as the maximum improvement and maximum decline observed. We do not report the numbers of performed iterations, as our focus is on the quality of the models and error-free responses were obtained in the first iteration in most cases. 

\begin{table*}[!t] 
\centering 
\caption{Impact of LLM self-optimization of output on model quality.} \label{tab:res:self-imp;output} 
\resizebox{\textwidth}{!}{ 
    \begin{tabular}{lcccccc} 
        \toprule 
        \multirow{2}{*}{\textbf{LLM}} & \textbf{Avg. Quality Before} & \textbf{Avg. Quality After} & \multirow{2}{*}{\textbf{Max. Improvement}} & \multirow{2}{*}{\textbf{Max. Decline}} \\
        & \textbf{Self-Improvement} & \textbf{Self-Improvement} & & \\
        \midrule
        \geminipro & 0.87 & 0.87 & +0.14 & -0.04 \\
        \geminiflash & 0.73 & 0.76 & +0.29 & -0.08 \\
        \gptfouro & 0.76 & 0.81 & +0.84 & -0.03 \\
        \bottomrule 
    \end{tabular} 
} 
\end{table*}

The results highlight that while the average improvements may appear modest, self-optimization of output can yield significant benefits in specific instances. \gptfouro\ showed the most substantial benefit, with the highest average quality gain (+0.05) and an instance of very large improvement (+0.84). \geminiflash\ also saw significant enhancements, achieving gains up to +0.29. For \geminipro, the impact of self-optimization was relatively low, with a slight average increase (+0.005), reflecting the high quality of its initial outputs. We also note that in some cases, self-optimization of output led to declines in quality. However, the maximum declines were relatively smaller compared to the maximum improvements, suggesting that while there is a risk of degradation, the potential for significant enhancement is greater.

In conclusion, our analysis indicates that allowing LLMs to self-optimize their outputs can, in general, be beneficial within our framework, especially for models that initially produce lower-quality outputs. However, there is a risk of quality degradation and it is crucial to design the prompt carefully to discourage hallucinations.


\section{Conclusion}
\label{sec:conclusion}
In this paper, we extended our LLM-driven process modeling framework by introducing a comprehensive benchmark and exploring LLM self-improvement strategies. Our evaluation of 16 state-of-the-art LLMs revealed substantial performance variations, with \claude\ demonstrating exceptional capabilities in generating high-quality process models efficiently. We found a positive correlation between error-handling performance and the overall quality of the generated models. Additionally, our analysis indicated consistent quality trends within similar model families.

The investigation of LLM self-improvement strategies revealed that while self-evaluation depends heavily on the chosen LLM and input optimization shows limited reliability, output optimization demonstrates promising potential for enhancing quality. This underscores the possibility of leveraging LLMs to autonomously refine their outputs, potentially reducing the need for manual intervention. However, carefully crafted prompts are crucial to discourage hallucinations.

Our work contributes valuable insights into the application of LLMs for automated process modeling. The benchmark provides a foundation for comparing LLM performance in this domain, while the self-improvement analysis identifies promising avenues for further enhancing LLM-generated process models. Future research directions include incorporating additional process perspectives beyond control-flow, exploring direct BPMN generation without intermediate representations, investigating alternative prompting strategies, and exploring the integration of external knowledge sources to further enhance the accuracy and reliability of LLM-generated process models.

\bibliography{references}

\end{document}